\documentclass[]{aa}

\usepackage{graphicx}
\usepackage{txfonts}

\usepackage{color}
\definecolor{gray}{RGB}{150,150,150}

\newcommand{\degree}{$^{\circ}~$}

\begin{document}
   \title{Cometary topography and phase darkening}

   \author{J.-B. Vincent
          \inst{1}
          }

   \institute{DLR Institute for Planetary Research\\
              Rutherfordstarsse 2, 12489, Berlin, Germany\\
              \email{jean-baptiste.vincent@dlr.de}
             }

   \date{Received December 06, 2018; accepted January 18, 2019}

%-------------------------------------------------------------------
% ABSTRACT
%-------------------------------------------------------------------
  \abstract
  % context heading (optional), leave it empty if necessary
  {}
   {Cometary surfaces can change significantly and rapidly due to the sublimation of their volatile material. Many authors have investigated this evolution; \cite{vincent2017} have used topographic data from all comets visited by spacecrafts to derive a quantitative model which relates large scale roughness (i.e. topography) with the evolution state of the nucleus for Jupiter Family Comets (JFCs). Meanwhile, ground based observers have published measurements of the phase functions of many JFCs and reported a trend in the phase darkening, with primitive objects showing a stronger darkening than evolved ones).}
   {In this paper, we use a numerical implementation of the topographic description by \cite{vincent2017} to build virtual comets and measure the phase darkening induced by the different levels of macro-roughness. We then compare our model with the values published by \cite{kokotanekova2018}}
   {We find that pure geometric effects like self-shadowing can represent up to 22\% of the darkening observed for more primitive objects, and 15\% for evolved surfaces. This shows that although physical and chemical properties remain the major contributor to the phase darkening, the additional effect of the topography cannot be neglected.}
  {}


   \maketitle
%

%-------------------------------------------------------------------
% INTRODUCTION
%-------------------------------------------------------------------
\section{Introduction}
Determining the age of a cometary surface is challenging. As the sublimation of volatile material ejects gas and refractory elements from the surface, most of this material is lost to space. It is estimated that the surface a typical Jupiter Family Comet (JFC) such as 9P/Tempel 1 or 67P/Churyumov-Gerasimenko loses on average a couple of meters per orbit (e.g. \cite{paetzold2016} for 67P). In details, this loss is mostly concentrated in specific areas of the nucleus where tens of meters of material can be removed, while other remain unchanged (e.g. \cite{thomas2007, elmaarry2017} for 9P and 67P, respectively). Most of the physical processes involved in this erosional activity remain to be understood, especially concerning the time scale at which they occur. Hence the difficulty in accurately measuring how evolved a nucleus is.
Yet, this is a critical measurement we need to obtain in order to understand the Solar System evolution. As comets are thought to be remnants of the early accretion, proto-planetesimals which never became larger, one must understand how much of the surface we observe now is really pristine and represents these initial conditions. Are the morphological and chemical features we observe primitive ? Or do they reflect more the evolutionary processes at work ?

Many authors have discussed cometary evolution, and a lot of progress has been made thanks to ESA's Rosetta mission which followed comet 67P from 2014 to 2016, through its perihelion passage. Detailed information about the evolutionary erosion observed on this comet can be found in \cite{elmaarry2017, birch2017, vincent2017, vincent2018} and references therein.

Of course, 67P is not the only comet visited by a spacecraft, and the morphological features of six nuclei have been measured so far, twice in the case of comet 9P which was observed by both NASA's missions Deep Impact \citep{ahearn2005} and Stardust/NEXT \citep{ahearn2011}. \cite{vincent2017} combined the data from these observations and extracted a statistical description of the topography of cometary nuclei.
They propose that JFCs' surface, more exactly the cumulative size distribution of the height of topographic features, can be accurately described by power laws. The power slope of this distribution provides measure of the evolution state of the surface. The measured that primitive nuclei display a shallower power law (slope=-1.5) while evolved surfaces have a steeper distribution (slope=-2.3).
In effect, this means that primitive surfaces are characterized by a rough topography on a large scale, with deep pits and tall cliffs all over the nucleus, possibly a signature of the early collisional environment, or large outbursts during the first orbit in the inner Solar System. On the contrary, evolved nuclei are better described as being quite smooth, most features erased by the activity. This concept is summarized in Figure \ref{fig:evolution}, left panel.

While our results are consistent with all in-situ observations so far, and supported by modeling of the thermal processes at play \citep{keller2015b, vincent2017}, we are limited by the fact that only a few comets have been observed with enough accuracy to measure the power law mentioned above. And we cannot currently constrain the timescale of this resurfacing due to large uncertainty in the orbital evolution of JFCs beyond their last close encounter with Jupiter (see for instance the discussion in \cite{ip2016}).

Yet, there are thousands of additional observations available in the ground-based observer community and it should be possible to link some of their measured quantities (albedo, spectra, phase function, ...) with evolution models. In a recent paper, \cite{kokotanekova2018} measured the phase function of fourteen JFCs, ten of them not yet visited by space missions. They report on a remarkable trend in the slope of the phase function, which shows an increase in phase darkening for objects considered the most primitive (based on the current understanding of their orbital evolution). In other words, primitive comets appear relatively darker at large phase angle.

As these comets supposedly share similar composition and physical properties, \citep{kokotanekova2018} argue that they may be observing a signature of the erosional processes described in \cite{vincent2017}: primitive surfaces, with their tall cliffs and deep pits, will display far more shadows at high phase than smooth, eroded nuclei. Thus the fore mentioned higher phase darkening.

Here, we test this hypotheses numerically, with the aim to derive a law connecting evolution and phase darkening.

%-------------------------------------------------------------------
% METHODS
%-------------------------------------------------------------------
\section{Methods}
\subsection{Procedural generation of comets}
Previously, \cite{shepard1998} investigated the photometric effects of roughness on on planetary surfaces at the smallest scale at which shadows can be produced, using a fractal description of the surface.
In this work, we consider the larges scale roughness of the surface, that is morphological features clearly identified and catalogued by space missions, and which typically span distances of ten to several hundred meters. Typically these are larger than a few tens of meters. Small scale variation of roughness is also an important parameter to be studied, and readers are referred to \cite{thomas2018} for the regional distribution of micro-roughness on comet 67P,  \cite{marshall2018} for its effects on Rosetta measurements in visible, infrared, and microwaves, and \cite{longobardo2017} on how photometry can be an indicator of comets' surface roughness at small scales.

To evaluate the effect of topography on the phase darkening, we must first define this topography and specify their spatial distribution, lateral, and vertical extent, as a function of the comet age. Fortunately, the topography of comets at scales larger than 10~m is relatively easy to describe and dominated by either smooth plains or rough terrains scarred with pits and cliffs. The latter are mostly associated to partial or fully formed pits on all comets, except for a few retreating scarps at the edge of smooth regions on 9P or 67P (e.g. \citet{thomas2013, thomas2015, groussin2015a}).
In order to build this topography procedurally with a computer, we need to define the number of pits per cometary nucleus, their spatial distribution, their diameter and depth. We achieve that from the following assumptions:
\begin{itemize}
    \item The initial number of pits is calibrated from observations. Considering objects like 67P or  81P as typical for 5~km diameter primitive nuclei \citep{birch2017, vincent2017}, we measured about 20 cliffs (or pit walls) of 300~m height, which we use has a reference point for our distribution. Larger cliffs are observed but their numbers are not well constrained; the gravitational field on small bodies can change a lot across short distances, and what looks like a large wall may not be vertical all over its surface (see the discussion in \cite{vincent2017}).
    \item The number of cliffs of any size is given by the power laws in \cite{vincent2017}. We use the power slopes -1.5 for primitive surfaces and -2.3 for evolved ones.
    \item Depth and diameter of pits are correlated, \cite{vincent2015a} and \cite{ip2016} measured a depth-to-diameter ratio of $0.73 \pm 0.08$ for the most recently formed features, down to an impact-crater like ratio of 0.2 for more evolved pits on JFCs.
    \item Observations of pits on 81P \citep{brownlee2004} and 67P \citep{vincent2015a} show that their morphology is almost perfectly cylindrical, rather than conic.
\end{itemize}

From these assumptions, we generate a distribution of pits and create a virtual comet by carving out cylinders with these calculated dimensions from an initial 5~km sphere. This is done with a Python script running in the free and open source 3D creation software suite \textit{Blender} (https://www.blender.org/). The code and one example file are available as supplementary material.

\begin{figure*}[h!]
\centering
   \includegraphics[width=14cm]{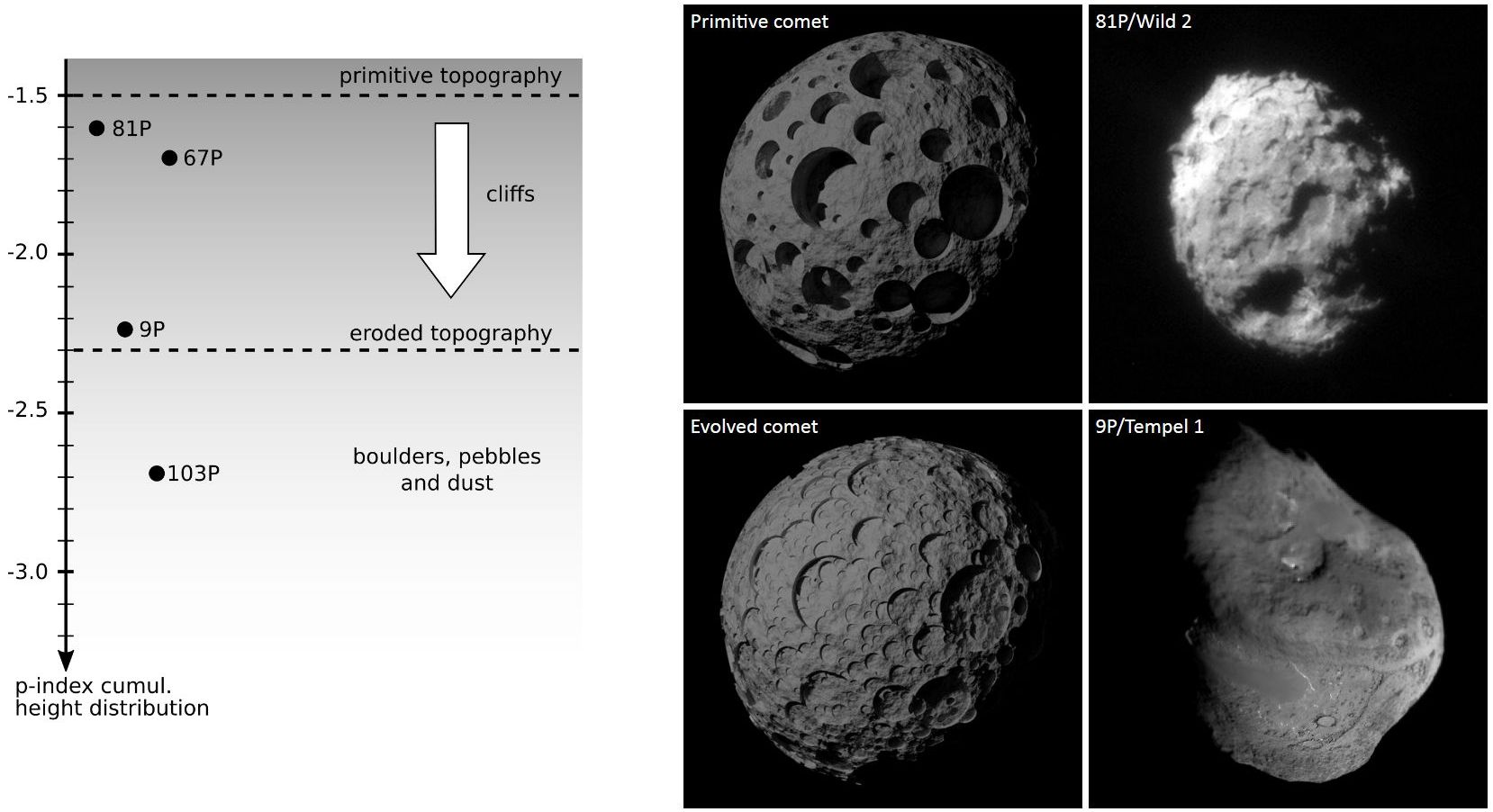}
    \caption{Left: Model of cometary evolution proposed by \cite{vincent2017}. Right: simulated primitive and evolved topography, compared to real nuclei assumed to belong to these categories.}
\label{fig:evolution}
\end{figure*}

Examples of comets generated with this technique can be seen in Figure \ref{fig:evolution}, right panel.

\subsection{Area phase function}
Positioning the observer and the Sun along the X-axis of the shape model thus generated, we rotate the Comet-Sun vector around the Z axis, from 0 to 110 degrees of phase, in steps of 10 degrees. At larger phases, when the Sun goes beyond the object horizon, the visible surface is in shadows and the phase function is controlled by the forward scattering properties of the material rather than the topography.

\begin{figure*}
\centering
   \includegraphics[width=14cm]{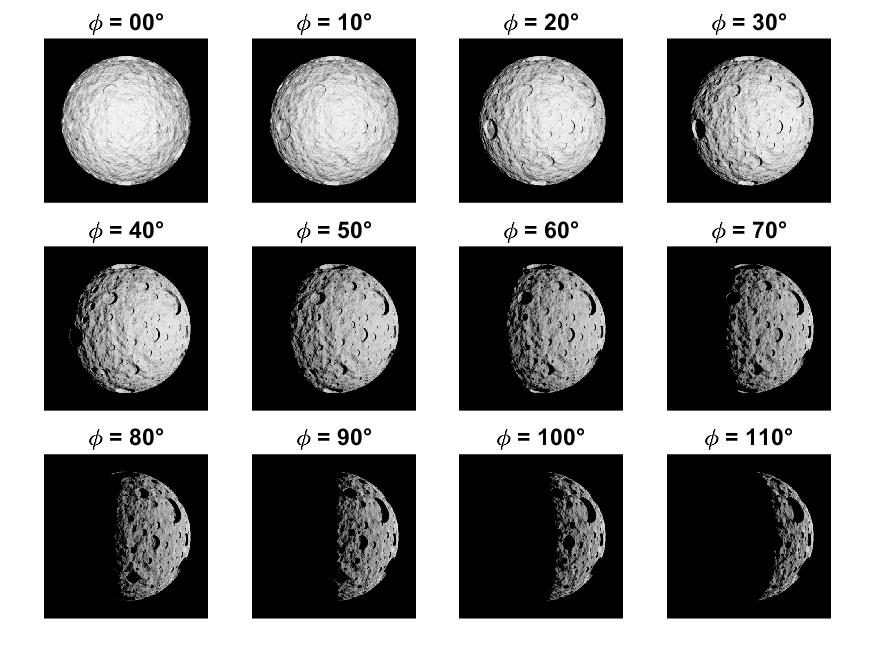}
    \caption{"Evolved" virtual comet at various phase angles. The default Lambertian bidirectional scattering distribution function of Blender is applied to this model in the final render for aesthetic reasons only. Our calculations only consider whether a pixel is lit or in shadow, regardless of the photometric function being used.}
\label{fig:phase_imgs}
\end{figure*}

For each phase angle, we raytrace the shadows cast by the topography and report how much of the illuminated surface is visible to the observer. This gives us a measure of the geometric phase darkening. An example of the varying illumination conditions is shown in Figure \ref{fig:phase_imgs}.

Its important to note that as we attempt to determine the contribution of topography to the phase function, we ignore all other parameters which could affect the photometry, like chemical and physical properties (e.g. albedo variations across the surface, grain size, refractory-to-ice ratio, ...).

Because the albedo of dark asteroids or comets does not change much across the spectral range of most of the solar energy, one can relate absolute magnitude $H$, geometric albedo $\alpha$, and diameter $D$ of an object with the simple relation \citep{harris1997}:
\begin{equation}\label{eq:1}
    log_{10}(D)= 3.1236 - 0.5 log_{10}(\alpha) - 0.2H
\end{equation}
with $D$ expressed in kilometers.\\

For each simulated phase simulated, we measure the illuminated and visible surface area, convert it to an equivalent disk diameter ($D = \sqrt{4 \times area / \pi}$) and derive its absolute magnitude by rewriting Equation \ref{eq:1} as:
\begin{equation}\label{eq:2}
    H = 5 \times [3.1236 - 0.5 log_{10}(\alpha)-log_{10}(D)];
\end{equation}

This gives us the photometric "area law" of our comets, defining what an observer would measure if the comet brightness was only a function of its illuminated surface \citep{lester1979}.

It is important to note that this law only describes the reduction of visible illuminated surface area, and does not intend to replace a full photometric model like \cite{hapke1993}. We argue that on airless bodies (including objects like comets where coma density close to the surface is extremely low), shadowed areas do not contribute to the overall brightness of the surface. It is true that some light scattering from nearby illuminated surfaces can allow one to resolve surface elements in shadows by stretching images acquired by spacecraft. However, the signal from those areas is close to the noise level and cannot be used for photometric measurements. It can be completely neglected for ground based observations.

%-------------------------------------------------------------------
% RESULTS
%-------------------------------------------------------------------
\section{Results}
\begin{figure}
  \resizebox{\hsize}{!}{\includegraphics{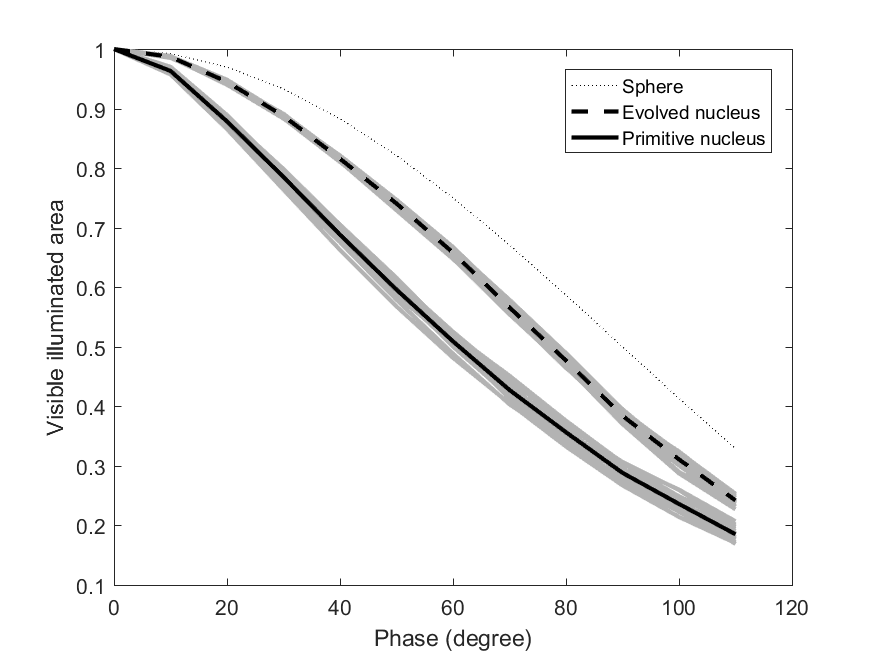}}
  \caption{Fraction of visible illuminated area for primitive and evolved comets as defined in the text, compared to a smooth sphere. Gray lines are all samples, 20 of each type, and black lines are the mean values. The standard deviation is $\pm 2\%$.}
  \label{fig:area_plot}
\end{figure}

We use the numerical approach described above to calculate the phase function of 40 procedurally generated comets (20 "primitive" and 20 "evolved") and display the average results in Figure \ref{fig:area_plot}.

As expected, the illuminated area is significantly reduced by increasing the large scale roughness of the topography. This effect is the largest for phase angles between 60\degree and 70\degree, for which the illuminated visible area fraction of a primitive surface is $15\% (\pm 2)$ smaller than that of an  evolved surface, and $25\% (\pm 2)$ smaller than for a smooth sphere. These results are easily understood qualitatively as an increase in topographic variations leads naturally to an increase in length of shadows at large phase angles, effectively masking large regions of the surface. For that same reason, north and south faces of mountains on Earth experience vastly different amount of daylight, which leads to remarkable ecological differences between both sides.

Incidentally, this effect has proven to be quite an issue for the planning of Rosetta observations at comet 67P, because even the subsolar point was not always illuminated due to the very rough topography in some areas (see morphological description in \citet{thomas2015}).\\

\begin{figure}
  \resizebox{\hsize}{!}{\includegraphics{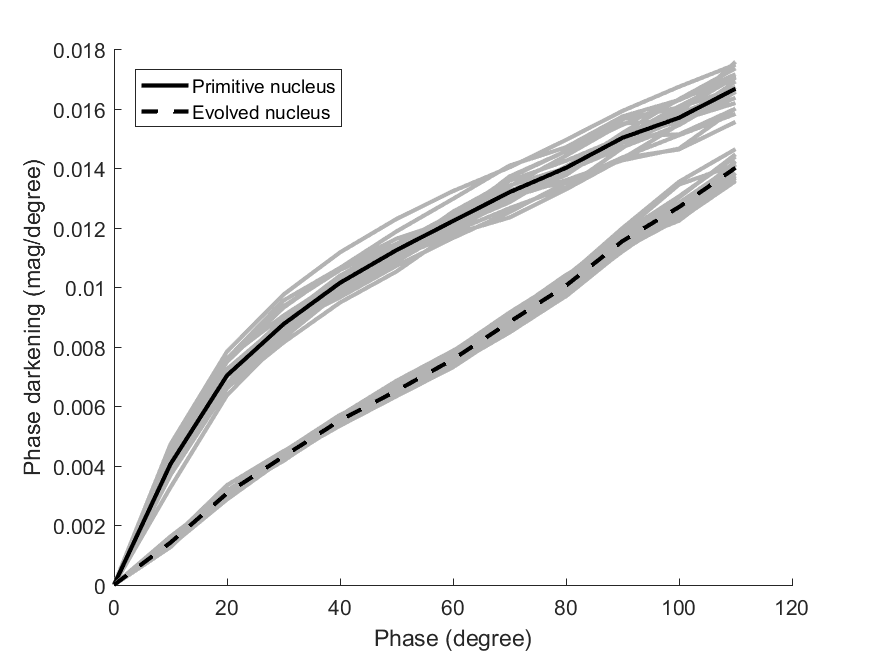}}
  \caption{Phase darkening as expected from self shadowing due to topographic features on primitive and evolved cometary nuclei.}
  \label{fig:mag_plot}
\end{figure}

After converting illuminated area to absolute magnitude, we derive a phase darkening which can be compared with observations by \cite{kokotanekova2018}. Our results are plotted in Figure \ref{fig:mag_plot}.
We find that the self shadowing created by the topography can result in a significant darkening of the surface. We note that this phase darkening is not linear, and increases with the phase angle. Across the whole range, our model predicts a mean darkening of 0.011 $\pm$0.001 mag/degree for primitive/rough objects and 0.007 $\pm 0.0005$ mag/degree for evolved/smooth ones, considering typical JFCs (diameter=5km, albedo=5\%).

We stress that this darkening is only due to the amount of shadows created by the topography. It should be interpreted as a correction factor to derive the effective surface contributing to the observed brightness. After this correction is applied, one still needs to consider a full photometric model in order to derive meaningful physical properties of the surface material.

Note that this geometric phase darkening is the derivative of Equation \ref{eq:2} with respect to the phase function, and as such it does not depend on the albedo of the surface (a constant measured at zero phase). Neither do we expect any dependence on the initial diameter if we assume that the power law describing the topography does not depend on the size of the object for comets of similar evolution status, which seems to be true for nuclei observed so far. Therefore, the space in between our two curves in Figure \ref{fig:mag_plot} should encompass all cometary nuclei for which our topographic description is valid.
The phase darkening appears to become quasi linear for evolved objects. This is because our smoothest modeled objects are close to spherical, and their visible illuminated surface area can be analytically described as $(1+cos(phase))/2$. This function is quasi-linear between  about 60\degree and 120\degree, an effect which is emphasized when we apply a square root and logarithm functions to proceed from surface area to magnitude.

When compared with the measurements published by \cite{kokotanekova2018}, we find that self-shadowing could explain 22\% of the phase darkening observed on a primitive object like 81P/Wild 2 and 15\% for evolved objects like 9P/Tempel 1.
The two values proposed here correspond to the roughest and smoothest cometary topographies observed by spacecrafts so far (respectively 81P or 67P, and 9P). Even smoother surfaces (e.g. 103P) barely show any topography and the effect described here is irrelevant. On the other extreme, it is possible to consider even rougher topographies but this is highly speculative as we have no evidence that such extreme terrains could be formed.

While not responsible for all darkening, topography is definitely a component which cannot be ignored. This puts constraints on the advanced photometric models (e.g.\cite{hapke1993}), which should not consider anymore the full phase darkening when trying to derive micro-roughness of the material but a reduced value corrected for large scale topographic effects by our model.

%-------------------------------------------------------------------
% CONCLUSIONS
%-------------------------------------------------------------------
\section{Conclusions}
From a statistical analysis of the distribution of large-scale topographic features, \cite{vincent2017} proposed a model of cometary surface evolution which can predict how cometary nuclei may look like, as a function of their evolution status.

\cite{kokotanekova2018} and coauthors have observed significant variations in the amount of phase darkening observed for Jupiter Family Comets, and suggest that it may also reflect how evolved those objects are.

In this follow-up work, we use our topographic model to generate virtual comets at different levels of evolution and calculate their phase darkening. Our numerical experiment suggests that topography can play a significant role in controlling the amount of observable illuminated surface area, and the resulting brightness.

We find that pure geometric effects like self-shadowing can represent up to 22\% of the darkening observed for more primitive objects, and 15\% for evolved surfaces. This shows that although physical and chemical properties remain the major contributor to the phase darkening, the additional effect of the topography cannot be neglected.

The idea that objects showing the strongest darkening are also the most primitive \citep{kokotanekova2018} is consistent with the evolution model proposed by \citep{vincent2017}.

\begin{acknowledgements}
      This research has made use of NASA's Astrophysics Data System.
\end{acknowledgements}

\bibliographystyle{aa} % style aa.bst
\bibliography{AA_2018_34789_accepted} % your references Yourfile.bib

\begin{thebibliography}{23}
\expandafter\ifx\csname natexlab\endcsname\relax\def\natexlab#1{#1}\fi

\bibitem[{{A'Hearn} {et~al.}(2011){A'Hearn}, {Belton}, {Delamere}, {Feaga},
  {Hampton}, {Kissel}, {Klaasen}, {McFadden}, {Meech}, {Melosh}, {Schultz},
  {Sunshine}, {Thomas}, {Veverka}, {Wellnitz}, {Yeomans}, {Besse}, {Bodewits},
  {Bowling}, {Carcich}, {Collins}, {Farnham}, {Groussin}, {Hermalyn}, {Kelley},
  {Kelley}, {Li}, {Lindler}, {Lisse}, {McLaughlin}, {Merlin}, {Protopapa},
  {Richardson}, \& {Williams}}]{ahearn2011}
{A'Hearn}, M.~F., {Belton}, M.~J.~S., {Delamere}, W.~A., {et~al.} 2011,
  Science, 332, 1396

\bibitem[{{A'Hearn} {et~al.}(2005){A'Hearn}, {Belton}, {Delamere}, {Kissel},
  {Klaasen}, {McFadden}, {Meech}, {Melosh}, {Schultz}, {Sunshine}, {Thomas},
  {Veverka}, {Yeomans}, {Baca}, {Busko}, {Crockett}, {Collins}, {Desnoyer},
  {Eberhardy}, {Ernst}, {Farnham}, {Feaga}, {Groussin}, {Hampton}, {Ipatov},
  {Li}, {Lindler}, {Lisse}, {Mastrodemos}, {Owen}, {Richardson}, {Wellnitz}, \&
  {White}}]{ahearn2005}
{A'Hearn}, M.~F., {Belton}, M.~J.~S., {Delamere}, W.~A., {et~al.} 2005,
  Science, 310, 258

\bibitem[{{Birch} {et~al.}(2017){Birch}, {Tang}, {Hayes}, {Kirk}, {Bodewits},
  {Campins}, {Fernandez}, {de Freitas Bart}, {Kutsop}, {Sierks}, {Soderblom},
  {Squyres}, \& {Vincent}}]{birch2017}
{Birch}, S.~P.~D., {Tang}, Y., {Hayes}, A.~G., {et~al.} 2017, \mnras, 469, S50

\bibitem[{{Brownlee} {et~al.}(2004){Brownlee}, {Horz}, {Newburn}, {Zolensky},
  {Duxbury}, {Sandford}, {Sekanina}, {Tsou}, {Hanner}, {Clark}, {Green}, \&
  {Kissel}}]{brownlee2004}
{Brownlee}, D.~E., {Horz}, F., {Newburn}, R.~L., {et~al.} 2004, Science, 304,
  1764

\bibitem[{{El-Maarry} {et~al.}(2017){El-Maarry}, {Groussin}, {Thomas},
  {Pajola}, {Auger}, {Davidsson}, {Hu}, {Hviid}, {Knollenberg}, {G{\"u}ttler},
  {Tubiana}, {Fornasier}, {Feller}, {Hasselmann}, {Vincent}, {Sierks},
  {Barbieri}, {Lamy}, {Rodrigo}, {Koschny}, {Keller}, {Rickman}, {A'Hearn},
  {Barucci}, {Bertaux}, {Bertini}, {Besse}, {Bodewits}, {Cremonese}, {Da
  Deppo}, {Debei}, {De Cecco}, {Deller}, {Deshapriya}, {Fulle}, {Gutierrez},
  {Hofmann}, {Ip}, {Jorda}, {Kovacs}, {Kramm}, {K{\"u}hrt}, {K{\"u}ppers},
  {Lara}, {Lazzarin}, {Lin}, {Lopez Moreno}, {Marchi}, {Marzari}, {Mottola},
  {Naletto}, {Oklay}, {Pommerol}, {Preusker}, {Scholten}, \&
  {Shi}}]{elmaarry2017}
{El-Maarry}, M.~R., {Groussin}, O., {Thomas}, N., {et~al.} 2017, Science, 355,
  1392

\bibitem[{{Groussin}(2015)}]{groussin2015a}
{Groussin}, O. e.~a. 2015, \aap

\bibitem[{{Hapke}(1993)}]{hapke1993}
{Hapke}, B. 1993, {Theory of reflectance and emittance spectroscopy} (Cambridge
  University Press)

\bibitem[{{Harris} \& {Harris}(1997)}]{harris1997}
{Harris}, A.~W. \& {Harris}, A.~W. 1997, \icarus, 126, 450

\bibitem[{{Ip} {et~al.}(2016){Ip}, {Lai}, {Lee}, {Cheng}, {Li}, {Lin},
  {Vincent}, {Besse}, {Sierks}, {Barbieri}, {Lamy}, {Rodrigo}, {Koschny},
  {Rickman}, {Keller}, {Agarwal}, {A'Hearn}, {Barucci}, {Bertaux}, {Bertini},
  {Bodewits}, {Boudreault}, {Cremonese}, {Da Deppo}, {Davidsson}, {Debei}, {De
  Cecco}, {El-Maarry}, {Fornasier}, {Fulle}, {Groussin}, {Guti{\'e}rrez},
  {G{\"u}ttler}, {Hviid}, {Jorda}, {Knollenberg}, {Kovacs}, {Kramm},
  {K{\"u}hrt}, {K{\"u}ppers}, {La Forgia}, {Lara}, {Lazzarin},
  {L{\'o}pez-Moreno}, {Lowry}, {Marchi}, {Marzari}, {Michalik}, {Mottola},
  {Naletto}, {Oklay}, {Pajola}, {Thomas}, {Toth}, \& {Tubiana}}]{ip2016}
{Ip}, W.-H., {Lai}, I.-L., {Lee}, J.-C., {et~al.} 2016, \aap, 591, A132

\bibitem[{{Keller} {et~al.}(2015){Keller}, {Mottola}, {Davidsson},
  {Schr{\"o}der}, {Skorov}, {K{\"u}hrt}, {Groussin}, {Pajola}, {Hviid},
  {Preusker}, {Scholten}, {A'Hearn}, {Sierks}, {Barbieri}, {Lamy}, {Rodrigo},
  {Koschny}, {Rickman}, {Barucci}, {Bertaux}, {Bertini}, {Cremonese}, {Da
  Deppo}, {Debei}, {De Cecco}, {Fornasier}, {Fulle}, {Guti{\'e}rrez}, {Ip},
  {Jorda}, {Knollenberg}, {Kramm}, {K{\"u}ppers}, {Lara}, {Lazzarin}, {Lopez
  Moreno}, {Marzari}, {Michalik}, {Naletto}, {Sabau}, {Thomas}, {Vincent},
  {Wenzel}, {Agarwal}, {G{\"u}ttler}, {Oklay}, \& {Tubiana}}]{keller2015b}
{Keller}, H.~U., {Mottola}, S., {Davidsson}, B., {et~al.} 2015, \aap, 583, A34

\bibitem[{{Kokotanekova} {et~al.}(2018){Kokotanekova}, {Snodgrass}, {Lacerda},
  {Green}, {Nikolov}, \& {Bonev}}]{kokotanekova2018}
{Kokotanekova}, R., {Snodgrass}, C., {Lacerda}, P., {et~al.} 2018, \mnras, 479,
  4665

\bibitem[{{Lester} {et~al.}(1979){Lester}, {McCall}, \& {Tatum}}]{lester1979}
{Lester}, T.~P., {McCall}, M.~L., \& {Tatum}, J.~B. 1979, \jrasc, 73, 233

\bibitem[{{Longobardo} {et~al.}(2017){Longobardo}, {Palomba}, {Capaccioni},
  {Ciarniello}, {Tosi}, {Mottola}, {Moroz}, {Filacchione}, {Raponi}, {Quirico},
  {Zinzi}, {Capria}, {Bockelee-Morvan}, {Erard}, {Leyrat}, {Rinaldi}, \&
  {Dirri}}]{longobardo2017}
{Longobardo}, A., {Palomba}, E., {Capaccioni}, F., {et~al.} 2017, \mnras, 469,
  S346

\bibitem[{{Marshall} {et~al.}(2018){Marshall}, {Groussin}, {Vincent}, {Brouet},
  {Kappel}, {Arnold}, {Capria}, {Filacchione}, {Hartogh}, {Hofstadter}, {Ip},
  {Jorda}, {K{\"u}hrt}, {Lellouch}, {Mottola}, {Rezac}, {Rodrigo}, {Rodionov},
  {Schloerb}, \& {Thomas}}]{marshall2018}
{Marshall}, D., {Groussin}, O., {Vincent}, J.-B., {et~al.} 2018, \aap, 616,
  A122

\bibitem[{{P{\"a}tzold} {et~al.}(2016){P{\"a}tzold}, {Andert}, {Hahn}, {Asmar},
  {Barriot}, {Bird}, {H{\"a}usler}, {Peter}, {Tellmann}, {Gr{\"u}n},
  {Weissman}, {Sierks}, {Jorda}, {Gaskell}, {Preusker}, \&
  {Scholten}}]{paetzold2016}
{P{\"a}tzold}, M., {Andert}, T., {Hahn}, M., {et~al.} 2016, \nat, 530, 63

\bibitem[{{Shepard} \& {Campbell}(1998)}]{shepard1998}
{Shepard}, M.~K. \& {Campbell}, B.~A. 1998, \icarus, 134, 279

\bibitem[{{Thomas} {et~al.}(2018){Thomas}, {El Maarry}, {Theologou},
  {Preusker}, {Scholten}, {Jorda}, {Hviid}, {Marschall}, {K{\"u}hrt},
  {Naletto}, {Sierks}, {Lamy}, {Rodrigo}, {Koschny}, {Davidsson}, {Barucci},
  {Bertaux}, {Bertini}, {Bodewits}, {Cremonese}, {Da Deppo}, {Debei}, {De
  Cecco}, {Fornasier}, {Fulle}, {Groussin}, {Guti{\`e}rrez}, {G{\"u}ttler},
  {Ip}, {Keller}, {Knollenberg}, {Lara}, {Lazzarin}, {L{\`o}pez-Moreno},
  {Marzari}, {Tubiana}, \& {Vincent}}]{thomas2018}
{Thomas}, N., {El Maarry}, M.~R., {Theologou}, P., {et~al.} 2018, \planss, 164,
  19

\bibitem[{{Thomas} {et~al.}(2015){Thomas}, {Sierks}, {Barbieri}, {Lamy},
  {Rodrigo}, {Rickman}, {Koschny}, {Keller}, {Agarwal}, {A'Hearn}, {Angrilli},
  {Auger}, {Barucci}, {Bertaux}, {Bertini}, {Besse}, {Bodewits}, {Cremonese},
  {Da Deppo}, {Davidsson}, {De Cecco}, {Debei}, {El-Maarry}, {Ferri},
  {Fornasier}, {Fulle}, {Giacomini}, {Groussin}, {Gutierrez}, {G{\"u}ttler},
  {Hviid}, {Ip}, {Jorda}, {Knollenberg}, {Kramm}, {K{\"u}hrt}, {K{\"u}ppers},
  {La Forgia}, {Lara}, {Lazzarin}, {Moreno}, {Magrin}, {Marchi}, {Marzari},
  {Massironi}, {Michalik}, {Moissl}, {Mottola}, {Naletto}, {Oklay}, {Pajola},
  {Pommerol}, {Preusker}, {Sabau}, {Scholten}, {Snodgrass}, {Tubiana},
  {Vincent}, \& {Wenzel}}]{thomas2015}
{Thomas}, N., {Sierks}, H., {Barbieri}, C., {et~al.} 2015, Science, 347,
  aaa0440

\bibitem[{{Thomas} {et~al.}(2007){Thomas}, {Veverka}, {Belton}, {Hidy},
  {A'Hearn}, {Farnham}, {Groussin}, {Li}, {McFadden}, {Sunshine}, {Wellnitz},
  {Lisse}, {Schultz}, {Meech}, \& {Delamere}}]{thomas2007}
{Thomas}, P.~C., {Veverka}, J., {Belton}, M.~J.~S., {et~al.} 2007, Icarus, 187,
  4

\bibitem[{{Thomas}(2013)}]{thomas2013}
{Thomas}, P. e.~a. 2013, Icarus, 222, 453

\bibitem[{{Vincent}(2018)}]{vincent2018}
{Vincent}, J.-B. 2018, in Lunar and Planetary Science Conference, Vol.~49,
  Lunar and Planetary Science Conference, 1280

\bibitem[{{Vincent} {et~al.}(2017){Vincent}, {Hviid}, {Mottola}, {Kuehrt},
  {Preusker}, {Scholten}, {Keller}, {Oklay}, {de Niem}, {Davidsson}, {Fulle},
  {Pajola}, {Hofmann}, {Hu}, {Rickman}, {Lin}, {Feller}, {Gicquel},
  {Boudreault}, {Sierks}, {Barbieri}, {Lamy}, {Rodrigo}, {Koschny}, {A'Hearn},
  {Barucci}, {Bertaux}, {Bertini}, {Cremonese}, {Da Deppo}, {Debei}, {De
  Cecco}, {Deller}, {Fornasier}, {Groussin}, {Guti{\'e}rrez},
  {Guti{\'e}rrez-Marquez}, {G{\"u}ttler}, {Ip}, {Jorda}, {Knollenberg},
  {Kovacs}, {Kramm}, {K{\"u}ppers}, {Lara}, {Lazzarin}, {Lopez Moreno},
  {Marzari}, {Naletto}, {Penasa}, {Shi}, {Thomas}, {Toth}, \&
  {Tubiana}}]{vincent2017}
{Vincent}, J.-B., {Hviid}, S.~F., {Mottola}, S., {et~al.} 2017, \mnras, 469,
  S329

\bibitem[{{Vincent} {et~al.}(2015){Vincent}, {Oklay}, {Marchi}, {H{\"o}fner},
  \& {Sierks}}]{vincent2015a}
{Vincent}, J.-B., {Oklay}, N., {Marchi}, S., {H{\"o}fner}, S., \& {Sierks}, H.
  2015, \planss, 107, 53

\end{thebibliography}

%-------------------------------------------------------------------
% ALL FIGURES
%-------------------------------------------------------------------

\end{document}